\documentclass[12pt]{article}
\def\be{\begin{equation}}
\def\ee{\end{equation}}
\def\bea{\begin{eqnarray}}
\def\eea{\end{eqnarray}}
\usepackage{graphicx}

\catcode`\@=11
\def\lsim{\mathrel{\mathpalette\@versim<}}
\def\gsim{\mathrel{\mathpalette\@versim>}}
\def\@versim#1#2{\vcenter{\offinterlineskip
\ialign{$\m@th#1\hfil##\hfil$\crcr#2\crcr\sim\crcr } }}
\catcode`\@=12
\usepackage{axodraw}

\catcode`@=12
\begin{document}
\thispagestyle{empty}
\begin{flushright}
UCRHEP-T454\\
KANAZAWA-08-07\\
September 2008\
\end{flushright}
\vspace{0.5in}
\begin{center}
{\LARGE \bf Fermion Triplet Dark Matter\\
and Radiative Neutrino Mass\\}
\vspace{1.0in}
{\bf Ernest Ma$^a$ and Daijiro Suematsu$^b$\\}
\vspace{0.2in}
{\sl $^a$ Department of Physics and Astronomy, University of California,\\ 
Riverside, California 92521, USA\\}
\vspace{0.1in}
{\sl $^b$ Institute for Theoretical Physics, Kanazawa University,\\  
920-1192 Kanazawa, Japan\\}
\end{center}
\vspace{1.0in}
\begin{abstract}\
The neutral member of a Majorana fermion triplet 
$(\Sigma^+,\Sigma^0,\Sigma^-)$ is proposed as a candidate for the dark 
matter of the Universe.  It may also serve as the seesaw anchor for 
obtaining a radiative neutrino mass.
\end{abstract}
   
\newpage
\baselineskip 24pt

\noindent \underline{\it Introduction}~: The cosmological and astrophysical 
evidence \cite{bhs05} for dark matter 
(DM) is a powerful incentive for considering new particles and interactions 
beyond those of the standard model (SM) of quarks and leptons.  Whereas 
most studies have concentrated on supersymmetric extensions of the SM, 
other excellent DM candidates abound.  For example, if the SM is extended 
to include just one new scalar or fermion multiplet, then there are many 
possible DM candidates \cite{cfs06}.  In particular, a scalar doublet 
$(\eta^+,\eta^0)$ odd under an exactly conserved $Z_2$ symmetry 
\cite{dm78} is a very good choice \cite{m06-1,bhr06,lnot07,glbe07}.

Such a ``dark'' scalar doublet is amenable to discovery at the Large 
Hadron Collider (LHC) \cite{cmr07}.  It is also very useful for generating 
small radiative Majorana neutrino masses \cite{m06-1} if there exist neutral 
singlet fermions $N_i$ which are odd under $Z_2$.  For a brief review of the 
further developments of this idea of ``scotogenic'' neutrino mass, see 
Ref.~\cite{m07-1}.  More recently, it has been extended to include 
$A_4$ tribimaximal mixing \cite{m08-1} as well.
 
Now the lightest $N_i$ may also be considered a DM candidate 
\cite{kst03,kms06,akrsz08}.  However, processes such as $\mu \to e \gamma$ 
impose severe constraints on the Yukawa couplings of $N_i$, making 
it difficult to satisfy the cosmological relic abundance required. 
One way to avoid this problem is to introduce additional interactions 
for $N_i$ \cite{ks06,bm08,s08}.  Other SM singlets have also been 
considered \cite{sz85,hllt07,blmrs08,m08-3,hln08,l08,ckk08,ks08,cms08}.

Whereas the canonical seesaw mechanism uses the fermion singlet $N$ so that 
the neutrino mass is given by $m_\nu \simeq -m_D^2/m_N$ where $m_D$ is the 
Dirac mass linking $\nu$ to $N$, it is not the only way to realize the 
generic dimension-five effective operator \cite{w79}
\begin{equation}
{\cal L}_5 = - {f_{ij} \over 2 \Lambda} (\nu_i \phi^0 - l_i \phi^+)
(\nu_j \phi^0 - l_j \phi^+) + H.c.
\end{equation}
for obtaining small Majorana neutrino masses in the SM.  In fact, there are 
three tree-level (and three generic one-loop) realizations \cite{m98}. The 
second most often considered mechanism for neutrino mass is that of a 
scalar triplet $(\xi^{++},\xi^+,\xi^0)$, whereas the third tree-level 
realization, i.e. that of a fermion triplet $(\Sigma^+,\Sigma^0,\Sigma^-)$ 
\cite{flhj89}, has not received as much attention.  However, it has some 
rather intriguing properties.  It supports a new U(1) gauge symmetry 
\cite{m02,mr02,bd05} and may be important for gauge-coupling unification 
\cite{m05,bs07,df07} in the SM.  It may be probed \cite{mr02,bns07,fhs08,aa08} 
at the LHC, and is being discussed in a variety of other contexts 
\cite{f07,abbgh08,ff08,m08-4,moy08}.  Now suppose $\Sigma^0$ is also odd 
under $Z_2$, then it may become a DM candidate \cite{m05,ff08} and replace $N$ 
in the radiative generation of neutrino mass as shown in Fig.~1.  
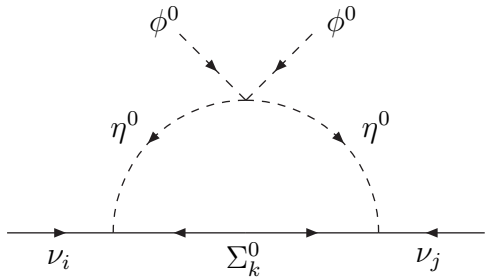
\begin{figure}[htb]
\begin{center}
\begin{picture}(360,120)(0,0)
\ArrowLine(90,10)(130,10)
\ArrowLine(180,10)(130,10)
\ArrowLine(180,10)(230,10)
\ArrowLine(270,10)(230,10)
\DashArrowLine(155,85)(180,60)3
\DashArrowLine(205,85)(180,60)3
\DashArrowArc(180,10)(50,90,180)3
\DashArrowArcn(180,10)(50,90,0)3
\Text(110,0)[]{$\nu_i$}
\Text(250,0)[]{$\nu_j$}
\Text(180,0)[]{$\Sigma^0_k$}
\Text(135,50)[]{$\eta^0$}
\Text(230,50)[]{$\eta^0$}
\Text(150,90)[]{$\phi^{0}$}
\Text(217,90)[]{$\phi^{0}$}
\end{picture}
\end{center}
\caption{One-loop generation of seesaw neutrino mass.}
\end{figure}
The difference between $N$ and $\Sigma^0$ is that whereas the former has 
only Yukawa interactions in the minimal scenario, the latter has electroweak 
gauge interactions, i.e. $\Sigma^0 \Sigma^\pm W^\mp$, which will allow 
$\Sigma^0$ and $\Sigma^\pm$ to annihilate and coannihilate in the early 
Universe to account for the correct DM relic abundance without relying 
on their Yukawa couplings \cite{kms06}.  Note that $\Sigma^\pm$ is slightly 
heavier than $\Sigma^0$ from electroweak radiative corrections \cite{cfs06}.   
It is also possible \cite{m08-2} that $\Sigma^0$ exists as DM without 
having anything to do with neutrino mass.

\noindent \underline{\it Gauge-coupling unification}~: It is well-known that 
gauge-coupling unification occurs for the minimal supersymmetric standard 
model (MSSM) but not the SM.  On the other hand, the addition of $\Sigma$ 
improves the situation and gauge-coupling unification in the SM is possible 
\cite{m05,bs07,df07} with the inclusion of some other fields.  Consider the 
one-loop renormalization-group equations governing the evolution of the 
three gauge couplings of the standard $SU(3)_C \times SU(2)_L \times U(1)_Y$ 
gauge group as functions of mass scale:
\begin{equation}
{1 \over \alpha_i(M_1)} - {1 \over \alpha_i(M_2)} = {b_i \over 2 \pi} \ln 
{M_2 \over M_1},
\end{equation}
where $\alpha_i = g_i^2/4 \pi$ and the numbers $b_i$ are determined by the 
particle content of the model between $M_1$ and $M_2$.  In the SM with one 
Higgs scalar doublet, these are given by
\begin{eqnarray}
SU(3)_C &:& b_C = -11 + (4/3) N_f = -7, \\ 
SU(2)_L &:& b_L = -22/3 + (4/3) N_F + 1/6 = -19/6, \\ 
U(1)_Y &:& b_Y = (4/3)N_f + 1/10 = 41/10,
\end{eqnarray}
where $N_f = 3$ is the number of quark and lepton families and $b_Y$ has been 
normalized by the well-known factor of 3/5.  Using the input \cite{pdg06}
\begin{eqnarray}
\alpha_L(M_Z) &=& (\sqrt{2}/\pi)G_F M^2_W = 0.0340, \\ 
\alpha_Y(M_Z) &=& \alpha_L(M_Z) \tan^2 \theta_W = 0.0102, \\ 
\alpha_C(M_Z) &=& 0.122,
\end{eqnarray}
it is easy to check that the SM particle content is incompatible with 
the unification condition
\begin{equation}
\alpha_C(M_U) = \alpha_L(M_U) = (5/3) \alpha_Y (M_U) = \alpha_U.
\end{equation}
Suppose $(\Sigma^+,\Sigma^0,\Sigma^-) \sim (1,3,0)$ and $(\eta^+,\eta^0) 
\sim (1,2,1/2)$ are added at the scale $M_X$, together with two real scalar 
color octets $\zeta_{1,2} \sim (8,1,0)$, then $\Delta b_L = 2(2/3) + 1/6 = 
3/2$, $\Delta b_Y = 1/10$, and $\Delta b_C = 3(2)(1/6) = 1$ between $M_X$ 
and $M_U$, so that Eq.~(9) implies
\begin{equation}
\ln {M_U \over M_Z} = \left( {\pi \over 45} \right) \left( 
{3 \over \alpha_Y(M_Z)} + {9 \over \alpha_L(M_Z)} - {14 \over \alpha_C(M_Z)} 
\right) = 31.0.
\end{equation}
Hence $M_U \simeq 2.65 \times 10^{15}~{\rm GeV}$, which is an acceptable value 
\cite{bmm82} for suppressing the proton decay lifetime above the experimental 
lower bound of about $10^{32}$ years.  The scale $M_X$ is determined to 
be about 730 GeV.  Thus the new particles have a chance of being observed at 
the LHC.  In particular, the $\zeta$ scalars would be produced in abundance 
at the LHC because they are color octets \cite{bm84, gw07} and would 
decay in one loop to two gluons \cite{m05}, i.e. $\zeta \to \zeta \zeta 
\to g g$.

\noindent \underline{\it $\Sigma^0$ as dark matter}~:  Consider the 
\underline{minimal} case where the SM is extended to include only one 
fermion triplet $\Sigma = (\Sigma^+,\Sigma^0,\Sigma^-) \sim (1,3,0)$ which 
is odd under $Z_2$ with all other fields even.  In that case, $m_{\Sigma^\pm} 
= m_{\Sigma^0}$ at tree level, but the former is heavier than the latter 
from one-loop electroweak radiative corrections, namely \cite{cfs06}
\begin{equation}
\Delta = m_{\Sigma^\pm} - m_{\Sigma^0} = {\alpha_L m_\Sigma \over 4 \pi} \left\{ 
f\left( {M_W \over m_\Sigma} \right) - \cos^2 \theta_W f\left( {M_Z \over 
m_\Sigma} \right) \right\},
\end{equation}
where
\begin{eqnarray}
f(r) &=& -r^2 + r^4 \ln r + r(r^2-4)^{1/2} (1+r^2/2) \ln [-1-(r^2-4)^{1/2}r/2 
+ r^2/2] \nonumber \\ &\simeq& 2 \pi r - 3 r^2, ~~{\rm for}~r 
\ll 1.
\end{eqnarray}
This splitting is positive and approaches $(\alpha_L/2)\cos \theta_W(1-
\cos \theta_W)M_Z \simeq 167$ MeV for large $m_\Sigma$.  This means that 
$\Sigma^\pm$ is allowed to decay into $\Sigma^0$ plus a virtual $W^\pm$ 
which then converts into $\pi^\pm$ or leptons.

The relic abundance of $\Sigma^0$ is determined by the annihilation 
and coannihilation of itself and $\Sigma^\pm$.  These cross sections are 
dominated by their $s$-wave contributions.  
For $\Sigma^0 \Sigma^0 \to W^+ W^-$ through $\Sigma^\pm$ exchange, 
\begin{equation}
\sigma (\Sigma^0\Sigma^0)|v| \simeq {2 \pi \alpha_L^2 \over m_\Sigma^2}, 
\end{equation}
where $v$ is the relative velocity of the incident particles in their center 
of mass and $m_\Sigma \gg \Delta$ is assumed.  As for coannihilation, several 
processes have to be included: $\Sigma^0 \Sigma^\pm \to W^0 W^\pm$ through 
$\Sigma^\pm$ exchange and $\Sigma^0 \Sigma^\pm \to W^\pm \to \bar{f} f',
~W^\pm W^0,~W^\pm H$, as well as $\Sigma^+ \Sigma^-\to W^0W^0$ through 
$\Sigma^\pm$ exchange, $\Sigma^+ \Sigma^- \to W^+W^-$ through 
$\Sigma^0$ exchange, $\Sigma^+ \Sigma^- \to W^0 \to \bar{f} f,
~W^+ W^-,~W^0 H$, and $\Sigma^\pm \Sigma^\pm \to W^\pm W^\pm$ through 
$\Sigma^0$ exchange.

They are also easily calculated to be
\begin{equation}
\sigma (\Sigma^0\Sigma^\pm)|v| \simeq {29 \pi \alpha_L^2 \over 8 m_\Sigma^2}, 
\qquad \sigma (\Sigma^+ \Sigma^-)|v| \simeq {37 \pi \alpha_L^2 
\over 8 m_\Sigma^2}, \qquad \sigma (\Sigma^\pm \Sigma^\pm)|v| \simeq 
{ \pi \alpha_L^2 \over m_\Sigma^2}.
\end{equation}
In the above, we have kept only the $a_{ij}$ coefficients in the 
relative-velocity expansion of the cross section:  $\sigma_{ij}|v| = 
a_{ij} + b_{ij}v^2$.  Note that $\sigma (\Sigma^0\Sigma^0)|v|$ is smaller 
than  $\sigma (\Sigma^0\Sigma^\pm)|v|$ and $\sigma (\Sigma^+\Sigma^-)|v|$.  
This means that $\Sigma^\pm$ contributes importantly to the relic 
abundance of $\Sigma^0$.

Using the method developed in Ref.~\cite{gs91} to take coannihilation into 
account, we calculate below the relic abundance of $\Sigma^0$ 
as a function of $m_\Sigma$ and $\Delta$.  The decoupling temperature $T_f$ 
of $\Sigma^0$ is estimated by using the effective cross section 
$\sigma_{\rm eff}$ and the effective degrees of freedom $g_{\rm eff}$ from 
the condition
\begin{equation}
x = \ln {0.038~g_{\rm eff}~M_{\rm Pl}~m_\Sigma~\langle\sigma_{\rm eff}|v|\rangle 
\over \sqrt{g_\ast x}},
\end{equation}
where $x=m_\Sigma/T$, $g_\ast = 106.75$ is the SM number of relativistic 
degrees of freedom at $T_f$, $M_{\rm Pl}=1.22\times 10^{19}$~GeV is the Planck 
mass, and
\begin{eqnarray}
\langle \sigma_{\rm eff}|v| \rangle &=& {g_0^2 \over g_{\rm eff}^2} 
\sigma (\Sigma^0\Sigma^0) + 4{g_0 g_\pm \over g_{\rm eff}^2} 
\sigma (\Sigma^0\Sigma^\pm) (1+\epsilon)^{3/2} \exp ({-\epsilon x}) 
\nonumber \\
&+& {g_{\pm}^2 \over g_{\rm eff}^2} [2 \sigma (\Sigma^+ \Sigma^-) + 2 \sigma 
(\Sigma^\pm \Sigma^\pm)](1+\epsilon)^2 \exp ({-2\epsilon x}), \nonumber \\
g_{\rm eff} &=& g_0+2 g_\pm(1+\epsilon)^{3/2} \exp ({-\epsilon x}),
\end{eqnarray}
with $g_0=g_\pm=2$ and $\epsilon = \Delta/m_\Sigma$.  
The relic abundance is then given by
\begin{equation}
\Omega h^2={1.04\times 10^9 x_f \over
g_\ast^{1/2}M_{\rm Pl}({\rm GeV})I_a},
\end{equation}  
where $I_a=x_f\int^\infty_{x_f} a_{\rm eff}x^{-2}dx$, $x_f = m_\Sigma/T_f$, 
and $a_{\rm eff}$ is extracted from  $\sigma_{\rm eff}|v|=a_{\rm eff}
+b_{\rm eff}v^2$.

Using the observational data $\Omega h^2=0.11\pm 0.006$ \cite{dm03}, we 
find $m_{\Sigma^0}$ to be in the range 2.28 to 2.42 TeV.  Here the 
electroweak radiative contribution to $\Delta$ is already at its 
asymptotic value of about 167 MeV and its effect on $m_{\Sigma^0}$ is 
negligible.  There is no tree-level contribution to $\Delta$ unless a 
Higgs triplet $(s^+,s^0,s^-)$ is added \cite{m05} with $\langle s^0 \rangle 
\neq 0$.  However, this value should be less than about 1 GeV to conform 
to precision electroweak measurements; hence $\Delta$ would still be 
negligible and our result is unchanged.

\noindent \underline{\it Neutrino masses}~:  To have scotogenic neutrino 
masses, consider now the addition of the dark scalar doublet $\eta$ and the 
specific choice of one $\Sigma$ and two $N$'s, then under the assumption 
$ m^2_\Sigma, m^2_N \ll m^2_\eta$, the resulting radiative masses are given by 
\cite{m06-1}
\begin{equation}
({\cal M}_\nu)_{\alpha\beta}={\lambda_5 v^2\over 8\pi^2}
\sum_{j=0,1,2}{h_{\alpha j}h_{\beta j}M_j \over m_\eta^2},
\label{mass}
\end{equation}
where $M_0=m_\Sigma$, $M_{1,2} = m_{N_{1,2}}$, $h_{\alpha j}$ are their 
Yukawa couplings, $v = \langle \phi^0 \rangle$, and $\lambda_5$ is the 
scalar coupling in the quartic term $(\lambda_5/2)(\Phi^\dagger \eta)^2 + 
H.c.$ which splits Re($\eta^0$) and Im($\eta^0$).  Since $\lambda_5$ and 
$m_\eta$ are adjustable, it is clear that realistic neutrino masses may be 
obtained for $h \sim 10^{-2}$, in which case processes such as $\mu \to 
e \gamma$ are well below their experimental upper bounds.  The problem 
with $N$ as dark matter is the requirement of $h > 1$ for it to have 
a large enough annihilation cross section \cite{kms06}.

Looking at the form of Eq.~(18), it is clear that it is possible to 
have a one-to-one correspondence betweeen the neutrino mass eigenvalues 
$m_{1,2,3}$ and the seesaw anchor masses $M_{0,1,2}$.  As an {\it anstaz}, 
let the $3 \times 3$ Yukawa coupling matrix linking $e,\mu, \tau$ to 
$M_{0,1,2}$ be given by
\begin{equation}
h_{\alpha j} = \pmatrix{\sqrt{2/3} & 1/\sqrt{3} & 0 \cr -1/\sqrt{6} & 
1/\sqrt{3} & -1/\sqrt{2} \cr -1/\sqrt{6} & 1/\sqrt{3} & 1/\sqrt{2}} 
\pmatrix{h_0 & 0 & 0 \cr 0 & h_1 & 0 \cr 0 & 0 & h_2},
\end{equation}
then the tribimaximal mixing of neutrinos is obtained, and their mass 
eigenvalues are
\begin{equation}
m_{i+1} = {\lambda_5 v^2 h_i^2 M_i \over 8 \pi^2 m_\eta^2},~~~i = 0,1,2.
\end{equation}

\noindent \underline{\it Relaxation of $\mu \to e \gamma$ constraints}~:
Since $\Sigma^0$ has gauge interactions, its relic abundance is adequately 
accounted for.  There is no need for it to have large Yukawa couplings, 
as is in the case \cite{kms06} of choosing the singlet fermion $N$ as 
dark matter, where $m_\eta$ must also be close to $m_N$.  This means 
radiative flavor-changing decays are easily suppressed.  In the above 
example, using the experimental upper bound of $1.2 \times 10^{-11}$ on 
the branching fraction of $\mu \to e \gamma$, this corresponds to the 
condition
\begin{equation}
| |h_0|^2 - |h_1|^2 | < 0.77 (m_\eta/2.35~{\rm TeV})^2.
\end{equation}
Since $h$ is not required to be large and $\eta$ should be heavier than 
$\Sigma$, the tension between the constraints of dark-matter relic abundance 
and flavor-changing radiative decays is removed.
   
\noindent \underline{\it Conclusion}~: In this paper we have proposed 
the addition of a fermion triplet $(\Sigma^+,\Sigma^0,\Sigma^-)$ to the 
standard model of quarks and leptons.  We consider $\Sigma^0$ as a 
dark-matter candidate, being odd under an exactly conserved $Z_2$ 
symmetry.  We show that with $\Sigma^\pm$ slightly heavier than $\Sigma^0$ 
from electroweak radiative corrections, $m_\Sigma^0 \sim 2.35$ TeV yields 
the correct dark-matter relic abundance from the annihilation and 
coannihilation of $\Sigma$ through gauge interactions.

We also consider $\Sigma$ as the seesaw anchor in the radiative generation 
of neutrino mass with a second scalar doublet $\eta$.  The constraints 
due to flavor-changing radiative decays such as $\mu \to e \gamma$ are 
then easily satisfied because the $\Sigma$ Yukawa couplings need not 
be large.  (If $\Sigma$ is replaced by $N$, then $N$ must have large 
Yukawa couplings to be a dark-matter candidate.)

E.M. was supported in part by the U.~S.~Department of Energy under Grant 
No.~DE-FG03-94ER40837.  D.S. was supported in part by a Grant-in-Aid for 
Scientific Research (C) from the Japan Society for Promotion of Science
(No.17540246).

\newpage
\bibliographystyle{unsrt}

\end{document}